\documentclass{article}

\usepackage{parskip}
\usepackage{graphicx}
\usepackage{amsmath, amssymb}
\usepackage{svg}
\svgpath{{images/}}
\usepackage{float}

\title{Lazy Diffusion: Mitigating spectral collapse in generative diffusion-based stable autoregressive emulation of turbulent flows}

\author{
  Anish Sambamurthy\\
  Department of Applied Mathematics\\
  University of California, Santa Cruz
  \and
  Ashesh Chattopadhyay\thanks{Corresponding author: \texttt{ashesh@ucsc.edu}}\\
  Department of Applied Mathematics\\
  University of California, Santa Cruz
}

\date{} 

\begin{document}

\maketitle

\section*{Abstract}
Turbulent flows posses broadband, power-law spectra in which multiscale interactions couple high-wavenumber fluctuations to large-scale dynamics. Although diffusion-based generative models offer a principled probabilistic forecasting framework, we show that standard DDPMs induce a fundamental \emph{spectral collapse}: a Fourier-space analysis of the forward SDE reveals a closed-form, mode-wise signal-to-noise ratio (SNR) that decays monotonically in wavenumber, $|k|$ for spectra $S(k)\!\propto\!|k|^{-\lambda}$, rendering high-wavenumber modes indistinguishable from noise and producing an intrinsic spectral bias. We reinterpret the noise schedule as a spectral regularizer and introduce power-law schedules $\beta(\tau)\!\propto\!\tau^\gamma$ that preserve fine-scale structure deeper into diffusion time, along with \emph{Lazy Diffusion}, a one-step distillation method that leverages the learned score geometry to bypass long reverse-time trajectories and prevent high-$k$ degradation. Applied to high-Reynolds-number 2D Kolmogorov turbulence and $1/12^\circ$ Gulf of Mexico ocean reanalysis, these methods resolve spectral collapse, stabilize long-horizon autoregression, and restore physically realistic inertial-range scaling. Together, they show that naïve Gaussian scheduling is structurally incompatible with power-law physics and that physics-aware diffusion processes can yield accurate, efficient, and fully probabilistic surrogates for multiscale dynamical systems.

\newpage
\section{Introduction}

Accurate emulation of turbulent dynamics is essential across the natural sciences and engineering, including aerospace and ocean engineering, combustion, and atmospheric and ocean modeling. While physics-based numerical simulations face prohibitive computational costs due to the need to resolve a broad range of spatiotemporal scales, data-driven surrogates encounter a fundamentally different barrier: the difficulty of reproducing multi-scale interactions and maintaining numerical stability over long forecasting horizons. In realistic settings, initial conditions observed from turbulent systems are noisy and often partial states can only be observed. This makes precise deterministic prediction infeasible beyond a few Lyapunov time scales. This limitation motivates a shift toward \textit{probabilistic forecasting}, where the objective is not to learn a point map $x_t \mapsto x_{t+\Delta t}$ but instead the conditional distribution $p(x_{t+\Delta t}\,|\,x_t)$ \cite{ehrendorfer1997predicting, buizza2008value,chattopadhyay2023long}. Such distributional models naturally account for uncertainty arising both from imperfect or noisy initial conditions and from model inadequacies introduced by the data-driven approximation, enabling the generation of diverse yet physically plausible trajectories.

Despite these conceptual advantages, applications of generative or otherwise probabilistic models to long-term turbulent emulation remain limited. Most existing deep learning surrogates target short-term dynamics \cite{kohl2023benchmarking, pathak2022fourcastnet,lam2022graphcast,bi2022pangu}, and extending them to long-range autoregressive prediction exposes structural instabilities. A central difficulty arises from \textit{spectral bias}: during training, gradients preferentially fit low-wavenumber structures, systematically under-representing high-wavenumber dynamics \cite{chattopadhyay2023challenges}. In systems governed by power-law energy spectra, such misrepresentation of high-wavenumber dynamics is consequential. High-wavenumber errors rapidly interact with and contaminate large-scale modes during autoregressive inference, triggering a progressive out-of-distribution (OOD) drift of the predicted states. As the predicted state departs from the training manifold, surrogates produce unphysical dynamics and ultimately become unstable \cite{pedersen2025thermalizer, chattopadhyay2023challenges,chattopadhyay2023oceannet}.

Recent hybrid approaches have attempted to address these issues. The G-LED framework \cite{gao2024generative} combines a transformer-based latent dynamical model with a diffusion model for subgrid-scale reconstruction, achieving strong performance on both the Kuramoto--Sivashinsky system and turbulent Navier--Stokes flows. The Thermalizer framework \cite{pedersen2025thermalizer} applies a U-Net for full-state autoregression while using a diffusion model trained on the original data distribution to resample states when predictions drift OOD. Our prior work, the FouRKS framework \cite{chattopadhyay2023challenges}, proposed stabilizing long-range deterministic emulation by enforcing spectral fidelity via Fourier-space losses during inference as a form of fine tuning. While effective individually, these approaches rely on composite architectures, auxiliary generative components, or explicit post-hoc correction mechanisms.

In this work, we pursue a fully \textit{generative diffusion-based}~\cite{ho2020denoising,sohl2015deep,song2021maximum} surrogate capable of stable autoregressive prediction for both high-Reynolds-number ($Re=10{,}000$) forced 2D turbulence and observed regional ocean dynamics. We show that spectral bias~\cite{rahaman2019spectral,chattopadhyay2023challenges,chattopadhyay2023oceannet,chakraborty2025binned}, well known in deterministic neural networks and operators, manifests in a distinct form in score-based generative models. Extending the spectral analysis of Chattopadhyay et al.~\cite{chattopadhyay2023challenges} to diffusion processes, we derive a theoretical characterization of \textit{signal-to-noise spectral collapse}: standard Gaussian noise schedules disproportionately obscure high-wavenumber dynamics early in the forward process, preventing the model from learning the fine-scale score components required for physically accurate long-term emulation.

Building on this understanding, we propose a two-stage strategy to mitigate spectral bias in diffusion models. First, using a continuous-time Stochastic Differential Equation (SDE) formulation, we reinterpret the noise schedule as a \textit{spectral regularization mechanism} and introduce \textit{power-law noise schedules} that preserve high-wavenumber coherence deep into the forward diffusion process. Second, we introduce \textit{Lazy Diffusion}, a novel one-step distillation scheme that leverages the learned score function to produce high-fidelity single-step generation, dramatically reducing inference cost while retaining spectral accuracy.

We validate these contributions on high Reynolds number Kolmogorov turbulence and high-resolution (8Km) regional ocean reanalysis data. Our results show that power-law-aware diffusion models substantially improve spectral fidelity, stabilize long-horizon emulation, and provide physically realistic forecasts. Together, these developments demonstrate that diffusion-based generative modeling---when appropriately adapted to multiscale physics---offers a promising path toward accurate, stable, and computationally efficient probabilistic surrogates for complex dynamical systems \cite{gilpin2024generative}.

\section{Systems and Datasets}
In this section, we introduce the two fully turbulent flow systems, \textbf{System}~1 (forced two-dimensional Kolmogorov turbulence) and \textbf{System}~2 (regional ocean surface reanalysis), from which all training, validation, and test data are derived.

\subsection{System 1}
We considered two systems. The first is the 2D forced Kolmogorov flow on a $\beta$-plane at $Re = 10{,}000$. The governing equations are
\[
\frac{\partial \omega}{\partial t} + \mathbf{u} \cdot \nabla \omega + \beta v = \frac{1}{Re} \nabla^2 \omega + \sin(4x) + \sin(4y),
\]
\[
\omega = \nabla^2 \psi,\quad \mathbf{u} = (-\partial_y \psi, \partial_x \psi),
\]
where $\beta = 20$ is the Coriolis parameter, $\omega$ is the vorticity, $\mathbf{u}$ is the velocity vector derived from the stream function $\psi$, and forcing has both zonal and meridional wavenumber $4$. We integrate this system with a doubly periodic pseudo-spectral solver at $256 \times 256$ resolution and $\Delta t_{DNS} = 5 \times 10^{-3}$ until steady-state turbulence is reached. From this simulation, we extract samples of $\omega(t)$ at intervals of $20\Delta t_{DNS}$. The vorticity fields are normalized to zero mean and unit variance for training.

\subsection{System 2}
The second system we analyze consists of surface ocean circulation dynamics over the Gulf of Mexico. Specifically, we use zonal surface velocity (SSU), and meridional surface velocity (SSV) fields extracted from the GLORYS global ocean reanalysis dataset~\cite{garric2018performance}, provided at $\frac{1}{12}^{\circ}$ horizontal resolution and daily temporal resolution. This reanalysis product combines satellite and in situ observations with the NEMO ocean circulation model~\cite{storkey2010forecasting} through a data assimilation framework. For training, we employ $7000$ temporal samples of SSU and SSV spanning the period from 1993 to 2012. Model evaluation is performed autoregressively, beginning from an initial state in 2013 and propagated forward for a duration of $1000$~days. These velocities are normalized to zero mean and unit variance.

\section{Methods}
\begin{figure}[H]
    \centering
    \includegraphics[width=\textwidth]{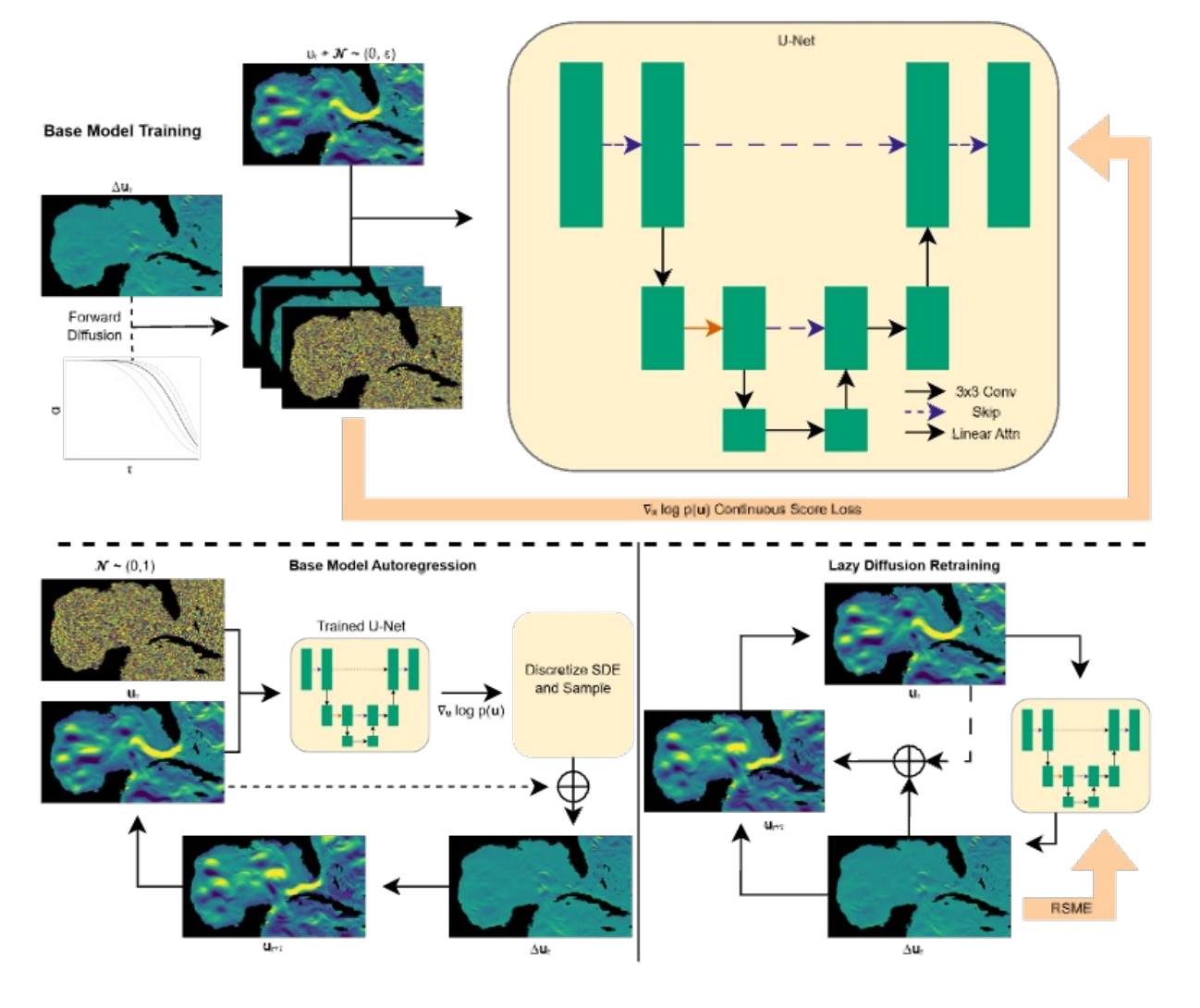}
    \caption{Schematic of the diffusion model and the lazy diffusion distillation.}
    \label{fig:schematic}
\end{figure}

Here, we detail the architectural design of our generative autoregressive models, the training protocols employed, and the proposed lazy-diffusion retraining procedure. Finally, we describe several additional engineering considerations and stabilization strategies that were essential for ensuring robust long-horizon autoregressive performance in the diffusion models. A schematic of the architecture design and other components is shown in Fig.~\ref{fig:schematic}.

\subsection{Score-Based Continuous Diffusion for Conditional Forecasting}

Our objective is to learn a conditional generative model for the one-step forecasting distribution
\[
p\!\left(X(t+\Delta t)\mid X(t)\right),
\]
which can be sampled autoregressively to produce long-horizon trajectories. To achieve this, we adopt the continuous score-based diffusion framework, in which a neural network learns the score of progressively noised data and uses it to integrate a reverse-time stochastic differential equation (SDE) that generates physically plausible future states conditioned on the current state.

\subsubsection{Forward Noising Process (Continuous-Time Limit of DDPM)}

Denoising Diffusion Probabilistic Models (DDPMs) introduce noise through a discrete Markov chain:
\[
x_\tau
=
\sqrt{1-\beta_\tau}\,x_{\tau-1}
+
\sqrt{\beta_\tau}\,\epsilon,
\qquad
\epsilon \sim \mathcal{N}(0,\mathbf{I}),
\]
with noise schedule $\{\beta_\tau\}_{\tau=1}^{\tau_f}$. This recursion has the closed form
\[
x_\tau
=
\sqrt{\bar{\alpha}_\tau}\,x_0
+
\sqrt{1-\bar{\alpha}_\tau}\,\epsilon,
\qquad
\bar{\alpha}_\tau = \prod_{s=1}^{\tau}(1-\beta_s),
\]
which clarifies that $\bar{\alpha}_\tau$ controls the remaining signal and $1-\bar{\alpha}_\tau$ controls the accumulated noise.

To obtain a differentiable-time formulation suitable for continuous score-based models, we examine the infinitesimal limit. Using the Taylor expansion
\[
\sqrt{1-\beta_\tau} = 1 - \tfrac{1}{2}\beta_\tau + O(\beta_\tau^2),
\]
the one-step increment becomes
\[
x_\tau - x_{\tau-1}
=
-\tfrac{1}{2}\beta_\tau\,x_{\tau-1}
+
\sqrt{\beta_\tau}\,\epsilon.
\]
Letting $\beta_\tau = \beta(\tau)\,d\tau$ and taking $d\tau \to 0$ yields the forward Itô SDE:
\[
dX
=
f(X,\tau)\,d\tau
+
g(\tau)\,dW_\tau,
\qquad
f(X,\tau) = -\tfrac{1}{2}\beta(\tau)X,
\quad
g(\tau)=\sqrt{\beta(\tau)}.
\]
This SDE defines a continuous noising process that maps data at $\tau = 0$ to an approximately Gaussian distribution at $\tau = \tau_f$, providing a mathematically coherent forward diffusion formulation.

\subsubsection{Training: Learning the Conditional Score}

Score-based diffusion models do not directly learn the conditional density
\[
p\!\left(X(t+\Delta t)\mid X(t)\right)
\]
itself. Instead, they learn the score of the intermediate noised distributions:
\[
\nabla_x \log p_\tau(x \mid X(t)).
\]
Using the forward SDE, one can sample $x_\tau$ in closed form via
\[
x_\tau
=
\sqrt{\bar{\alpha}_\tau}\,x_0
+
\sqrt{1-\bar{\alpha}_\tau}\,\epsilon,
\qquad
\epsilon \sim \mathcal{N}(0,I),
\]
where $x_0 = X(t+\Delta t)$ is the target future state conditioned on the known present state $X(t)$.

The score network $s_\theta(x,\tau, X(t))$ ($\theta$ are the parameters of the neural newtork) is trained using denoising score matching:
\[
\mathcal{L}(\theta)
=
\mathbb{E}_{\tau, x_0, \epsilon}
\left[
\left\|
s_\theta(x_\tau,\tau, X(t))
+
\frac{\epsilon}{\sqrt{1-\bar{\alpha}_\tau}}
\right\|_2^2
\right].
\]
This loss arises from the identity
\[
\nabla_x \log p_\tau(x_\tau \mid X(t))
=
-
\frac{\epsilon}{\sqrt{1-\bar{\alpha}_\tau}},
\]
and therefore encourages the network to estimate the true score of the noised conditional distribution at every diffusion time.

Thus, training consists of sampling a timestep $\tau$, generating $x_\tau$, and asking the model to predict the noise responsible for the corruption of $x_0$. The conditioning variable (the present state $X(t)$) is concatenated or encoded in the model input so that learning is performed over the entire conditional distribution.

\subsubsection{Sampling: Reverse-Time SDE for Conditional Generation}

Once the conditional score is learned, generation proceeds by integrating the reverse-time SDE corresponding to the forward diffusion. Anderson~\cite{anderson1982reverse} showed that for a forward SDE
\[
dX
=
f(X,\tau)d\tau + g(\tau)dW_\tau,
\]
the reverse process obeys
\[
dX
=
\left[
f(X,\tau)
- g(\tau)^2 \nabla_x \log p_\tau(x \mid X(t))
\right] d\tau
+
g(\tau)\,d\bar{W}_\tau,
\]
where $\bar{W}_\tau$ is a reverse-time Wiener process.

Replacing the true score with its neural approximation $s_\theta$ yields the practical sampling equation:
\[
dX
=
\left[
f(X,\tau)
- g(\tau)^2 s_\theta(X,\tau, X(t))
\right] d\tau
+
g(\tau)\,d\bar{W}_\tau.
\]
Integrating this SDE from $\tau=\tau_f$ back to $\tau=0$ transforms Gaussian noise into a sample from the learned conditional distribution:
\[
X(t+\Delta t) \sim p_\theta(\,\cdot \mid X(t)).
\]

Repeating this process autoregressively produces a full model rollout:
\[
X_{n+1} = \hat{X}(t_{n+1}) \sim p_\theta(\,\cdot \mid X_n),
\]
enabling long-horizon simulation of \textbf{System}~\textbf{1} and \textbf{System}~\textbf{2}.

\subsubsection{Power law noise schedulers}
\label{sec:power_law_noise}
As will be demonstrated in detail in section~\ref{sec:results}, the standard DDPM forward process when paired with a white-noise Gaussian corruption kernel, induces a rapid, wavenumber-dependent degradation of the signal–to–noise ratio, with high-wavenumber modes collapsing exponentially earlier than their low-wavenumber counterparts. Although more sophisticated kernels or adaptive scheduling mechanisms could potentially alleviate this effect, we focus on a simple and effective modification that is compatible with the continuous-time diffusion framework: a power-law noise schedule,
\[
\beta(\tau)
=
\beta_{\min}
+
(\beta_{\max}-\beta_{\min})\,\tau^{\gamma}.
\]
Increasing the exponent $\gamma$ concentrates the noise injection near the end of the forward diffusion trajectory. For example, increasing $\gamma$ from $1$ to $5$ shifts three-quarters of the total noise addition into the final $\sim 20\%$ of diffusion time, thereby delaying corruption of high-wavenumber components. This preserves fine-scale structure deeper into the forward process and provides the reverse-time sampler with more effective refinement capacity for the highest resolved modes.

To assess the influence of this scheduling parameter, we conduct a sweep over $\gamma \in [1.0, 7.0]$ and evaluate both the baseline score models and their lazily re-trained variants. As shown later in the Results section, power-law scheduling plays a critical role in mitigating spectral bias, improving high-wavenumber fidelity, and enhancing long-horizon stability in autoregressive emulation for both dynamical systems studied.

\subsection{Lazy Diffusion: Distilling the Score Model Into a Single-Step Conditional Predictor}
\label{sec:lazy_diff}

A key advantage of the score-based diffusion framework is that the network is trained on samples drawn from the entire forward diffusion trajectory. Thus, it learns the structure of the conditional data distribution at multiple corruption levels, enabling robust denoising across a wide range of noise intensities. For conditional forecasting, however, the eventual sampling objective is considerably simpler: given the present state $X(t)$, the model must generate a single draw from the conditional distribution
\[
p\!\left(X(t+\Delta t)\mid X(t)\right).
\]
This raises the question of whether multi-step numerical integration of the reverse-time SDE is necessary, or whether this mapping can be achieved directly.

\subsubsection{Limitations of Direct One-Step Denoisers}

Training a neural network to map a corrupted sample $x_{\tau^*}$ directly to the clean sample $x_0$ in one step is generally unstable. Without guidance from the diffusion process, the network must implicitly infer the geometry of the conditional data manifold and simultaneously remove noise and reconstruct fine-scale structure. In systems exhibiting broadband spectral content, such as turbulent flows and ocean circulation, this is a highly ill-posed regression problem prone to over-smoothing, spectral collapse, or mode failure.

\subsubsection{Insight From the Conditional Score}

The continuous-time diffusion formulation provides the structural information necessary to overcome these limitations. The pretrained score network 
\[
s_\theta(x,\tau, X(t)) \approx \nabla_x \log p_\tau(x \mid X(t))
\]
encodes the gradient of the log-density of the conditional distribution at each diffusion time $\tau$. This gradient characterizes the local geometry of the manifold of likely states (see section~\ref{sec:why_score_points}). For many diffusion formulations (e.g., variance-exploding SDEs), this score field directs corrupted samples toward regions of higher probability density, thereby providing a principled direction of denoising at each noise level.  

Crucially, this means that the pretrained score network already contains information describing how to transition from moderately corrupted samples toward the clean data distribution. The task of producing a one-step estimator can therefore be formulated not as learning this structure from scratch, but as distilling it into a direct mapping.

\subsubsection{Lazy Re-Training: A Distillation Procedure}

We propose \emph{lazy diffusion}, a second-stage training procedure that refines the pretrained score model into a single-step conditional predictor. Let $\tau^*$ denote an intermediate noise level at which the forward process yields a moderately corrupted sample. Using the closed-form forward diffusion process,
\[
x_{\tau^*}
=
\sqrt{\bar{\alpha}_{\tau^*}}\,x_0
+
\sqrt{1-\bar{\alpha}_{\tau^*}}\,\epsilon,
\qquad
\epsilon \sim \mathcal{N}(0,I),
\]
we construct supervised training pairs $(x_{\tau^*}, x_0)$.

A new neural network $F_\phi$ is initialized from the pretrained score model parameters and optimized to approximate the clean state via
\[
\mathcal{L}_{\text{lazy}}(\phi)
=
\mathbb{E}_{x_0,\epsilon}
\left[
\|F_\phi(x_{\tau^*}, X(t)) - x_0\|^2
\right].
\]
This distillation process does not require the model to infer the full reverse-time denoising trajectory. Instead, it leverages the fact that the score model already represents the local log-density geometry of the conditional distribution at $\tau^*$, allowing the distilled model to focus solely on aggregating this information into a single-step reconstruction of $x_0$.

\subsubsection{Sampling With Lazy Diffusion}

After lazy re-training, sampling from the conditional distribution reduces to two steps.  
First, we generate a corrupted sample at the chosen noise level $\tau^*$:
\[
x_{\tau^*}
=
\sqrt{\bar{\alpha}_{\tau^*}}\,x_0
+
\sqrt{1-\bar{\alpha}_{\tau^*}}\,\epsilon,
\qquad
\epsilon \sim \mathcal{N}(0,I).
\]
Second, we apply the distilled predictor to obtain the model’s estimate of the future state:
\[
\hat{X}(t+\Delta t)
=
F_\phi(x_{\tau^*}, X(t)).
\]
This produces a single-step approximation to the conditional forecast distribution without requiring numerical integration of the reverse SDE. In practice, lazy diffusion yields substantial computational savings while maintaining high-wavenumber fidelity and long-horizon stability in both \textbf{System}~\textbf{1} and \textbf{System}~\textbf{2}.

\subsection{Addressing Distribution Shift.}

A fundamental challenge in autoregressive emulation arises from the distribution shift that inevitably develops once model predictions are recursively fed back as inputs. During training, the model learns the conditional density
\[
p\!\left(X(t+\Delta t)\mid X(t)\right),
\]
where the conditioning state is always drawn from the true data manifold. During inference, however, the conditioning state rapidly departs from this manifold. Due to spectral bias in the learned dynamics, the earliest and largest errors occur at high wavenumbers; these errors then propagate into the conditioning input at the next timestep:
\[
p\!\left(X(t+\Delta t)\mid X(t)+\epsilon(t)\right),
\]
where \(\epsilon(t)\) denotes accumulated autoregressive prediction error. This mismatch places the model outside the distribution encountered during training, leading to accelerated error growth and eventual dynamical instability.

To mitigate this phenomenon, we introduce controlled perturbations to the conditioning states during training, chosen to match the magnitude and variability of prediction errors expected during autoregressive rollout. Consider a model prediction
\[
\hat{X}(t+\Delta t) = f_\theta\!\left(X(t)\right)
\]
with instantaneous error
\[
\epsilon(t)
=
\hat{X}(t+\Delta t) - X(t+\Delta t).
\]
If accumulated errors are modeled as isotropic Gaussian perturbations,
\[
\epsilon \sim \mathcal{N}(0, \sigma_\epsilon^2 I),
\]
then the mean squared error satisfies
\[
\mathrm{MSE}
=
\mathbb{E}\!\left[\|\hat{X}-X\|^2\right]
=
\sigma_\epsilon^2.
\]
A naive stabilization strategy would therefore set the conditioning noise standard deviation to
\(\sigma_\eta = \sqrt{\mathrm{MSE}}\).  
However, matching only the expected error amplitude is insufficient, since autoregressive degradation—particularly that produced by high-wavenumber attenuation—varies significantly across time and spatial scales.

Instead, we adopt a coverage-based criterion: conditioning noise should be chosen such that approximately half of the training perturbations lie within the expected error magnitude, and half exceed it, thereby improving robustness to out-of-distribution inputs. Formally, we require
\[
P\!\left(\|\eta\| \le \sigma_\epsilon\right) \approx 0.5,
\qquad
\eta \sim \mathcal{N}(0,\sigma_\eta^2 I).
\]
For one-dimensional marginals,
\[
P\!\left(|\eta_i| \le \sigma_\epsilon\right)
=
2\Phi\!\left(\frac{\sigma_\epsilon}{\sigma_\eta}\right)-1,
\]
and setting this probability to \(0.5\) yields
\[
\Phi\!\left(\frac{\sigma_\epsilon}{\sigma_\eta}\right) = 0.75
\quad\Longrightarrow\quad
\frac{\sigma_\epsilon}{\sigma_\eta}
=
\Phi^{-1}(0.75)
\approx 0.674,
\]
so that
\[
\sigma_\eta
\approx 1.48\, \sigma_\epsilon
\approx 1.5 \sqrt{\mathrm{MSE}}.
\]
This inflation factor ensures that conditioning samples span both the in-distribution regime (perturbations within expected predictive error) and a controlled degree of out-of-distribution deviation (perturbations exceeding the error scale), which is essential for stabilizing autoregressive trajectories.

During training, the perturbed conditioning input is thus constructed as
\[
X_t^{\mathrm{cond}}
=
X(t) + \eta,
\qquad
\eta \sim \mathcal{N}(0, \sigma_\eta^2 I),
\]
with \(\sigma_\eta\) determined from the coverage criterion above. This approach directly targets the distribution shift induced by spectral bias, improves robustness of the conditional model to off-manifold inputs, and significantly enhances long-horizon stability during autoregressive emulation.

\subsection{Network Architecture for the Conditional Score Model}

The conditional score network used throughout this study is based on a U-Net backbone augmented with sinusoidal diffusion-time embeddings and linear-time self-attention modules. This architecture is chosen for its ability to capture both local multi-scale structure and long-range spatial correlations, which are essential for accurately representing turbulent and oceanic dynamical fields.

\paragraph{General Structure.}
The network follows an encoder–decoder U-Net design with symmetric skip connections. Each resolution level contains two residual blocks, each consisting of a convolutional layer with Group Normalization and SiLU activation. Time information is injected at every block through a sinusoidal positional encoding passed through a two-layer MLP with a $4\times$ expansion of the hidden dimension. Self-attention layers are inserted at selected spatial resolutions to facilitate global information exchange without incurring prohibitive computational cost. The final projection head consists of
\[
\text{GroupNorm} \;\rightarrow\; \text{SiLU} \;\rightarrow\; \text{Conv}_{1\times 1}
\;\rightarrow\; \text{SiLU} \;\rightarrow\; \text{Conv}_{1\times 1},
\]
mapping the latent representation to the appropriate number of score channels.

\paragraph{System~1: Turbulent Flow.}
For the two-dimensional Kolmogorov turbulence dataset, we employ five resolution levels with channel multipliers $[2,\,4,\,4,\,8,\,16]$, applied to a base of 16 channels. This results in channel widths
\[
[16,\;32,\;64,\;64,\;128,\;256]
\]
across successive downsampling stages. Self-attention is applied at the $32\times 32$, $16\times 16$, and $8\times 8$ resolutions, which correspond to the scales most relevant for capturing the nonlinear energy cascade and the long-range coupling characteristic of turbulent flows. Each attention layer uses 32 heads with 64-dimensional head representations. Residual blocks employ Group Normalization with the number of groups
\[
\min\!\left(\frac{\text{channels}}{4},\,32\right),
\]
ensuring stable training across varying feature-map widths.

\paragraph{System~2: Ocean Reanalysis.}
The surface ocean circulation fields are defined on a non-square grid of $192\times 320$ with six input channels (two channels for predicted increments and four channels for two-timestep conditioning). To accommodate this geometry and maintain computational tractability, we use four resolution levels with channel multipliers $[2,\,4,\,4,\,8]$ applied to a 32-channel base width. Self-attention is introduced only at the $24\times 12$ and $12\times 6$ resolutions, where the spatial dimensionality is sufficiently reduced to allow efficient global mixing. The model predicts zonal and meridional surface velocities $(U,\,V)$, and a land–sea mask is applied during loss computation to exclude land points, ensuring that learning is focused exclusively on dynamically active ocean regions.

This architecture provides a unified backbone capable of representing the multi-scale, anisotropic structure of both turbulent flows and ocean circulation. The combination of hierarchical convolutional processing, scale-selective self-attention, and diffusion-time modulation enables the score model to capture fine-scale statistical structure while maintaining stability across the full range of noise levels used during training.



\subsection{Training and Evaluation Protocol}

\paragraph{System~1:}
Models for the turbulent flow dataset are trained using a batch size of 2 with the AdamW optimizer and a cosine learning-rate schedule decaying from $10^{-4}$ to $10^{-5}$ over 20{,}000 optimization steps. Gradient norms are clipped at $0.25$ to ensure numerical stability when learning high-wavenumber structure. Conditioning noise is introduced with standard deviation $\sigma_\eta = 0.01$, selected based on the single-step prediction error estimated from an initial pilot model. At each training iteration, four diffusion times are sampled independently from a uniform distribution,
\[
\tau \sim \mathcal{U}(0.001,\,1.0),
\]
which encourages learning of the score function across the full range of noise levels.  

\paragraph{System~2:}
For the ocean reanalysis dataset, we train using the $1/12^\circ$ GLORYS fields defined on a $192\times 320$ grid over the Gulf of Mexico, using data from 1993–2012. The model conditions on two preceding timesteps and predicts the increments $(\Delta U,\,\Delta V)$ for the zonal and meridional surface velocities. Training uses a batch size of 1 with the AdamW optimizer and cosine decay of the learning rate from $10^{-4}$ to $10^{-6}$ over 200{,}000 steps. Gradient clipping is applied at a threshold of $0.5$. Conditioning noise is set to $\sigma_\eta = 0.001$, derived from the error magnitude of the first autoregressive prediction step. Two diffusion times are sampled per iteration from
\[
\tau \sim \mathcal{U}(0.001,\,1.0),
\]
to span the full diffusion-time domain.

\paragraph{Lazy Diffusion Retraining.}
Lazy diffusion distillation is performed by fixing a single intermediate noise level, $\tau^* = 0.5$, and re-training the model for 20{,}000 additional iterations using the root-mean-squared error between the predicted and clean fields as the sole loss term. No auxiliary regularization or score-matching terms are employed. All other hyperparameters—including optimizer settings, gradient clipping, and conditioning noise—are inherited from the corresponding base score model. This stage converts the multi-step denoising process into a single-step conditional predictor while preserving high-wavenumber fidelity.

\paragraph{Inference and Evaluation.}
For score-based models, inference uses Euler–Maruyama integration of the reverse-time SDE with 1000 discretization steps, unless otherwise specified. Lazy diffusion models perform prediction in a single network evaluation. Autoregressive evaluation consists of 2000-step rollouts for System~1 and 1000-step rollouts for System~2. For Kolmogorov turbulence, spectral error is computed across all resolved wavenumbers $k$; for the ocean dataset, errors are evaluated across the corresponding zonal and meridional wavenumber bands. These long-horizon rollouts allow assessment of stability, spectral fidelity, and preservation of physically relevant statistics.





\section{Results}
\label{sec:results}
In this section, we first discuss the theoretical analysis of spectral collapse in diffusion models trained on turbulent energy spectra. We then provide a mitigative solution and implement it at scale for diffusion-based autoregressive emulation of \textbf{Systems} \textbf{1} and \textbf{2}.

\subsection{Theoretical spectral analysis of forward diffusion for turbulent energy spectra}

\label{sec: SNR_analysis}

A central property of both dynamical systems considered in this study, forced two-dimensional turbulence and regional ocean surface dynamics, is the presence of scale invariance within their inertial ranges. Empirically, this manifests as an isotropic power-law decay of the power spectral density (PSD),
\[
S_0(k) \;\propto\; |k|^{-\lambda}, \qquad \lambda > 0,
\]
for wavenumbers $k$ lying in the inertial subrange. This power-law structure has significant implications for the behavior of diffusion-based generative models when used for autoregressive emulation. In particular, the standard forward noising process in Denoising Diffusion Probabilistic Models (DDPMs) attenuates high-wavenumber dynamics more aggressively than low-wavenumber dynamics, leading to a wavenumber-dependent collapse of the signal-to-noise ratio (SNR), which we call spectral collapse in this paper. We now derive this phenomenon in detail.

Let $X_0 \in \mathbb{R}^d$ denote an initial physical state (e.g., a vorticity field), and consider the standard DDPM forward diffusion dynamics:
\begin{equation}
\label{eq:ddpm_forward}
X_\tau \;=\; \sqrt{\bar{\alpha}_\tau}\, X_0 
\;+\; \sqrt{1-\bar{\alpha}_\tau}\, \varepsilon,
\qquad
\varepsilon \sim \mathcal{N}(0,\sigma^2 I_d),
\end{equation}
where
\[
\bar{\alpha}_\tau = \prod_{s=1}^{\tau} \alpha_s, 
\qquad 
0 < \alpha_s < 1.
\]
This representation follows from the closed-form reparameterization of the multi-step Gaussian noising~\cite{DDPM}.

The scalar SNR at diffusion time $\tau$ is defined as
\begin{equation}
\label{eq:snr_scalar}
\mathrm{SNR}(\tau)
\;=\;
\dfrac{\mathbb{E}\left[\| \sqrt{\bar{\alpha}_\tau} X_0\|^2\right]}%
{\mathbb{E}\left[\| \sqrt{1-\bar{\alpha}_\tau}\, \varepsilon\|^2\right]}
\;=\;
\dfrac{\bar{\alpha}_\tau}{(1-\bar{\alpha}_\tau)\sigma^2},
\end{equation}
where we used $\mathbb{E}\|X_0\|^2 = \mathbb{E}\|\varepsilon\|^2 = d$ under unit variance normalization.

Let $\widehat{X}(k) = \mathcal{F}[X](k)$ denote the unitary discrete Fourier transform. Because the forward process, given by Eq.~\eqref{eq:ddpm_forward} is linear and convolution-free (additive white noise), it acts independently on each Fourier coefficient. Applying $\mathcal{F}$ to Eq.~\eqref{eq:ddpm_forward} gives:
\begin{equation}
\label{eq:fourier_forward}
\widehat{X}_\tau(k)
\;=\;
\sqrt{\bar{\alpha}_\tau}\,\widehat{X}_0(k)
\;+\;
\sqrt{1-\bar{\alpha}_\tau}\,\widehat{\varepsilon}(k),
\end{equation}
where independence and zero-mean of $X_0$ and $\varepsilon$ are mode-wise preserved.

Assume that the initial field has an isotropic power-law energy spectrum,
\begin{equation}
\label{eq:initial_spectrum}
S_0(k)
\;:=\;
\mathbb{E}\!\left[|\widehat{X}_0(k)|^2\right]
\;=\;
A\,|k|^{-\lambda},
\qquad 
A > 0,\; \lambda>0,
\end{equation}
consistent with turbulent inertial-range scaling. Moreover, since $\varepsilon$ is white Gaussian noise,
\[
\mathbb{E}\!\left[|\widehat{\varepsilon}(k)|^2\right] = \sigma^2
\qquad \text{for all } k.
\]

Taking the expectation of the squared magnitude of Eq.~\eqref{eq:fourier_forward}, and using the independence of $\widehat{X}_0(k)$ and $\widehat{\varepsilon}(k)$, yields:
\begin{align}
S_\tau(k)
&:=\;
\mathbb{E}\!\left[\,|\widehat{X}_\tau(k)|^2\,\right]
\\[0.2em]
&=\;
\bar{\alpha}_\tau\, 
\mathbb{E}\!\left[|\widehat{X}_0(k)|^2\right]
\;+\;
(1-\bar{\alpha}_\tau)\,
\mathbb{E}\!\left[|\widehat{\varepsilon}(k)|^2\right]
\\[0.3em]
&=\;
\bar{\alpha}_\tau\, A\,|k|^{-\lambda}
\;+\;
(1-\bar{\alpha}_\tau)\,\sigma^2.
\label{eq:Stau}
\end{align}
The cross-term vanishes because each component has zero mean and the pair is independent.

Equation \eqref{eq:Stau} shows explicitly that the noisy spectrum at time $\tau$ is a convex combination of a power-law signal and flat (white) noise.

We now compute the spectral SNR, which quantifies the relative strength of the physical signal and injected noise at each wavenumber:
\begin{equation}
\label{eq:snr_spectral}
\mathrm{SNR}(k,\tau)
\;:=\;
\dfrac{\bar{\alpha}_\tau A |k|^{-\lambda}}
{(1-\bar{\alpha}_\tau)\sigma^2}.
\end{equation}

For fixed $\tau$, this SNR decays as a power law in $|k|$. Differentiating Eq.~\eqref{eq:snr_spectral} with respect to $|k|$ gives
\begin{align}
\frac{\partial}{\partial |k|}
\mathrm{SNR}(k,\tau)
&=\;
\dfrac{\bar{\alpha}_\tau A}{(1-\bar{\alpha}_\tau)\sigma^2}\,
\frac{\partial}{\partial |k|}
|k|^{-\lambda}
\\[0.3em]
&=\;
-\,\dfrac{\bar{\alpha}_\tau A\,\lambda}
{(1-\bar{\alpha}_\tau)\sigma^2}\,
|k|^{-\lambda -1}.
\label{eq:snr_derivative}
\end{align}
Since all constants are positive and $|k|^{-\lambda-1}>0$, we obtain the strict inequality
\[
\frac{\partial}{\partial |k|}\mathrm{SNR}(k,\tau)
\;<\; 0,
\qquad \forall\, k\neq 0.
\]

Equation \eqref{eq:snr_spectral} combined with Eq.~\eqref{eq:snr_derivative} reveals a key limitation of standard diffusion noising: the spectral SNR decays polynomially in $|k|$, with exponent $\lambda > 0$ determined by the physics of the underlying turbulent cascade. Thus, high-wavenumber dynamics (large $|k|$) lose signal power far more rapidly than low-wavenumber components as $\tau$ increases. This is shown in both Fig.~\ref{fig:SNR} (a) and (b), where the lines corresponding to $\gamma=1$ (standard DDPM with linear noise schedule) for the high-wavenumber component of the signal decay far more quickly than those in mid wavenumbers or low wavenumbers. As discussed in our proposed power law noise schedulers in section~\ref{sec:power_law_noise}, increasing the value of $\gamma$ improves the preservation of the signal in the high wavenumbers late into the forward diffusion process (discussed further in section~\ref{sec:loss_analysis}). 

Once $\mathrm{SNR}(k,\tau)$ drops below the learnability threshold of the score model (parameterized by a U-Net in our case), the corresponding Fourier mode becomes effectively indistinguishable from noise. This occurs at much earlier diffusion times for high-wavenumber modes, causing the diffusion model to learn an increasingly biased representation of the data: large-scale structures remain visible throughout the diffusion horizon, while inertial-range and dissipation-range features are overwhelmed by noise.

This stratified decay demonstrates that the forward diffusion process erases information in a wavenumber-dependent manner: large-scale modes survive nearly intact, inertial-range modes degrade gradually, and dissipation-range modes vanish almost immediately. Because the score network is trained to denoise samples across all diffusion times, it receives virtually no usable supervision for high-$k$ modes at moderate or large $\tau$. As a result, the learned score field is intrinsically biased toward low-wavenumber reconstruction. Figure~\ref{fig:SNR}(a) makes this mechanism explicit: the diffusion process itself imposes a spectral preference that inevitably leads to high-wavenumber underfitting and long-term instability in autoregressive emulation.

This phenomenon is the root of the spectral bias in diffusion models when applied to turbulent systems with power-law spectra. It directly affects the model's ability to reconstruct fine-scale corrections during sampling and consequently limits the fidelity and stability of long-horizon autoregressive emulation.

\subsection{Biased loss weighting at the high wavenumbers}

\label{sec:loss_analysis}

\begin{figure}[H]
    \centering
    \includegraphics[width=\textwidth]
    {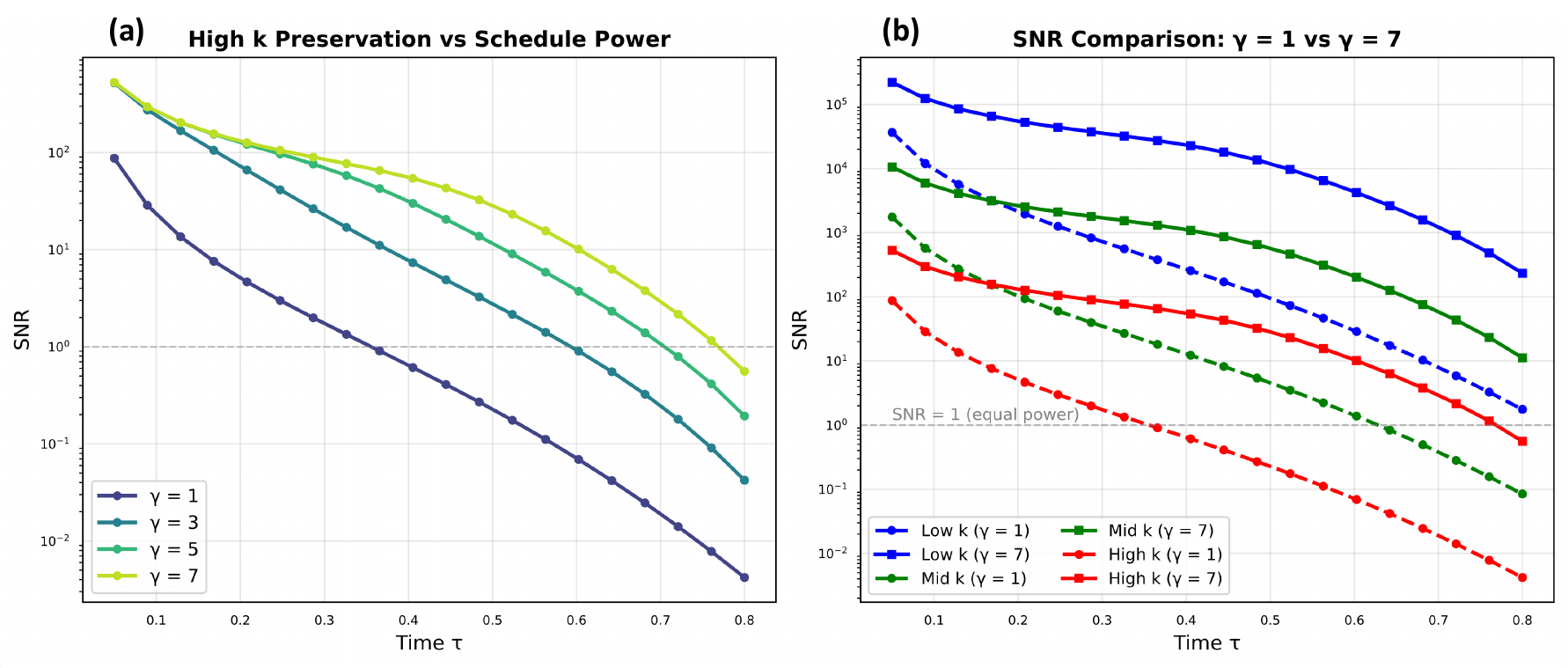}
    \caption{Signal to Noise Ratio for low, medium, and high k frequency bands for a power law adopting signal using different power values $\gamma$. SNR = 0 represents the white noise floor of the DDPM process.}
    \label{fig:SNR}
\end{figure}

The wavenumber-stratified collapse of $\mathrm{SNR}(k,\tau)$ shown in Fig.~\ref{fig:SNR}(a) and (b) reveals that, under a standard DDPM noise schedule, the forward diffusion process progressively eliminates learnable information in a manner that is fundamentally incompatible with the physics of turbulent flows. In particular, while low-wavenumber (large-scale) modes retain coherent signal until the final diffusion timestep, mid-$k$ inertial-range modes lose usable SNR halfway through the process, and high-$k$ modes collapse almost immediately. This implies that, for a large portion of the diffusion horizon, all high-wavenumber Fourier components that reach the score network during training are effectively indistinguishable from white noise.

This observation has a direct and profound consequence on the DDPM training objective. Recall that the score network is trained via the denoising objective
\[
\mathcal{L}(\theta)
=
\mathbb{E}_{\tau,X_0,\varepsilon}
\!\left[
\left\|
\mathbf{s}_\theta(X_\tau,\tau)
+
\frac{\varepsilon}{\sqrt{1-\bar{\alpha}_\tau}}
\right\|^2
\right].
\]
Because $X_\tau$ is distributed according to the forward process
$X_\tau = \sqrt{\bar{\alpha}_\tau} X_0 + \sqrt{1-\bar{\alpha}_\tau}\,\varepsilon$,
the target that the network must predict is precisely the noise field $\varepsilon$ rescaled by $(1-\bar{\alpha}_\tau)^{-1/2}$. At diffusion times where the SNR for a given wavenumber shell has collapsed to the noise floor (as is the case for high-$k$ modes at very small $\tau$), the score model is asked to fit a target that contains \emph{pure noise} in that shell. Thus, the score-matching loss forces the model to devote capacity to predicting an irreducible Gaussian random variable at high $k$, even though those modes in the true data distribution contain meaningful turbulent structure at $\tau=0$.

To see this relationship explicitly, we analyze the loss in spectral space. Parseval's theorem allows the pixel-space objective to be rewritten as an integral over the Fourier domain:
\[
\widehat{\mathcal{L}}(\theta)
\propto
\mathbb{E}_{\tau, X_0, \varepsilon}
\left[
\int_{\mathbb{R}^2}
\left|
\widehat{\mathbf{s}}_\theta(\mathbf{k},\tau)
+
\frac{\widehat{\varepsilon}(\mathbf{k})}{\sqrt{1-\bar{\alpha}_\tau}}
\right|^2
\, d\mathbf{k}
\right].
\]
In two spatial dimensions, the wavenumber measure becomes
$d\mathbf{k} = |k|\, d|k|\, d\phi$,
and for isotropic turbulence, the angular dependence integrates out. The loss reduces to
\[
\widehat{\mathcal{L}}(\theta)
\propto
\mathbb{E}_{\tau,X_0,\varepsilon}
\left[
\int_0^{k_{\max}}
\left|
\widehat{\mathbf{s}}_\theta(|k|,\tau)
+
\frac{\widehat{\varepsilon}(|k|)}{\sqrt{1-\bar{\alpha}_\tau}}
\right|^2
\, |k|\, d|k|
\right].
\]
The key geometric factor $|k|$ arises because the number of Fourier modes in a wavenumber shell $[k, k+dk]$ grows linearly with $k$. As a result, the contribution of each shell to the loss scales as the shell area in Fourier space.

Combining this observation with the turbulent power spectrum
\[
S_0(k) \propto |k|^{-\lambda},
\qquad \lambda > 1,
\]
reveals a deep structural misalignment between the DDPM objective and the physics of turbulence. The integrand for low-wavenumber shells decays as
$|k|^{1-\lambda}$,
which is integrable and small for realistic turbulent spectra. In contrast, the high-wavenumber contribution grows proportionally to $|k|$, since the noise spectrum is flat and the shell geometry dominates. Thus, the spectral form of the loss disproportionately weights precisely the Fourier bands that the forward process has already corrupted into white noise (high $k$), while assigning extremely small weight to the dynamically important, energy-containing large scales (low $k$).

The consequence is that the network effectively expends the majority of its representational capacity attempting to fit irreducible Gaussian noise in the high-wavenumber shells, rather than reconstructing the physically meaningful turbulent structures carried by the low-$k$ and mid-$k$ modes. This is the spectral analogue of the SNR collapse illustrated in Fig.~\ref{fig:SNR}(a) and (b): the forward process with a linear noise schedule destroys high-wavenumber signal long before the network ever sees it, and the DDPM loss then forces the network to waste capacity on predicting those already-destroyed modes. 


This wavenumber-dependent imbalance directly motivates the two mitigation strategies we develop in this work. First, we introduce \emph{power schedules}, which delay the injection of noise until very late in the diffusion trajectory. By keeping $\bar{\alpha}_\tau$ near unity for the majority of diffusion time, the SNR for all wavenumber bands, including the high-$k$ shells, remains above the noise floor for a significantly longer duration as shown in both Fig~\ref{fig:SNR}.(a) and (b) for power schedules, $\gamma >> 1$. This preserves high-wavenumber signal deeper into the forward process and provides the score network with meaningful supervision for those modes.

Second, we develop a \emph{lazy diffusion} distillation (see section~\ref{sec:lazy_diff}) procedure that bypasses most of the forward diffusion entirely. By exploiting the unimodal structure of the conditional distributions encountered in autoregressive prediction, we train a model to perform a single-shot denoising step. This approach sidesteps the early high-$k$ SNR collapse by eliminating the long diffusion trajectory in which high-wavenumber information is destroyed. Together, these two mitigation strategies directly counteract the structural biases introduced by the DDPM forward process and restore meaningful high-wavenumber learning essential for accurate turbulent emulation.

\subsection{Autoregressive emulation with power-law schedules and lazy diffusion}

Using the theoretical insights derived in sections~\ref{sec: SNR_analysis} and~\ref{sec:loss_analysis}, we now evaluate three autoregressive emulation strategies at scale using data from \textbf{Systems} \textbf{1} and \textbf{2}: (i) a standard DDPM with a linear noise schedule, (ii) a continuous score-based diffusion model using power-law noise schedules with exponents $\gamma \in [2.5, 7.0]$, and (iii) our proposed \emph{lazy diffusion}, a distilled single-step score model designed to bypass the instability of long diffusion trajectories.

\begin{figure}[H]
    \centering
    \includegraphics[width=\textwidth]{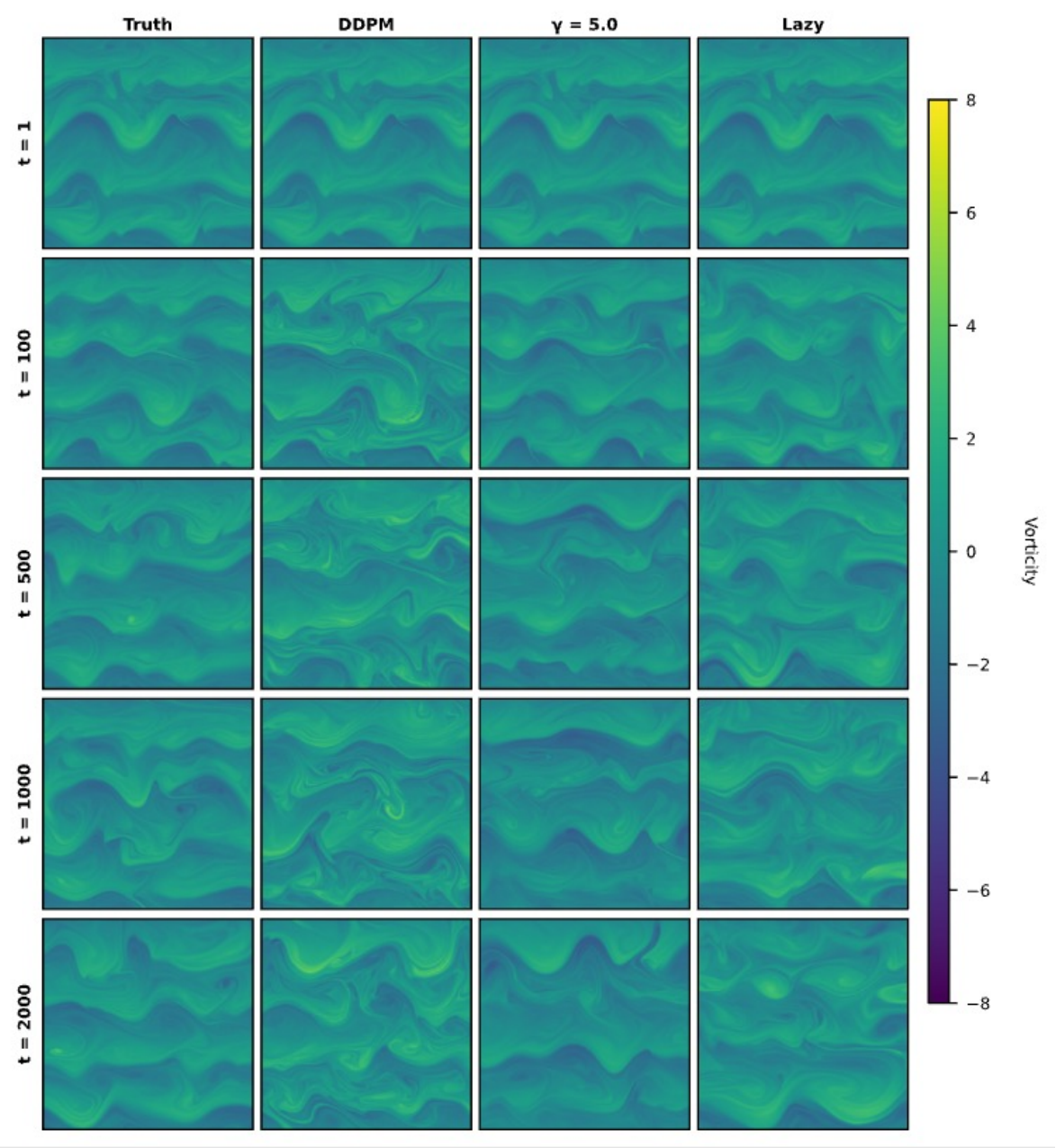}
    \caption{Autoregressive evolution of vorticity fields in \textbf{System~1}. Snapshots shown at $t = \{1, 100, 500, 1000, 2000\}$ timesteps ($t=20\Delta t_{DNS}$) for the ground truth, linear‐schedule DDPM, the power‐schedule model with $\gamma = 5.0$, and its lazy‐diffusion variant.}
    \label{fig:_Sys1_snapshot}
\end{figure}

In \textbf{System~1}, Fig.~\ref{fig:_Sys1_snapshot} shows that although all models generate superficially plausible turbulence at early times, their long‐term behavior diverges sharply. The ground truth maintains four coherent jets aligned with the meridional and zonal forcing at wavenumber~4, a hallmark feature of statistically stationary Kolmogorov flow. This jet structure is preserved only by the $\gamma = 5.0$ model across thousands of autoregressive steps. The linear-schedule DDPM instead exhibits periodic jet collapse and reformation, consistent with the theoretical prediction that early SNR collapse eliminates high-wavenumber structure and disrupts the energy balance between forcing and dissipation. The lazy diffusion model, despite using only a single denoising step, retains substantially more flow coherence than DDPM, reflecting the benefits of bypassing the high-$k$ signal destruction inherent to long forward-diffusion trajectories.

\begin{figure}[H]
    \centering
    \includegraphics[width=\textwidth]{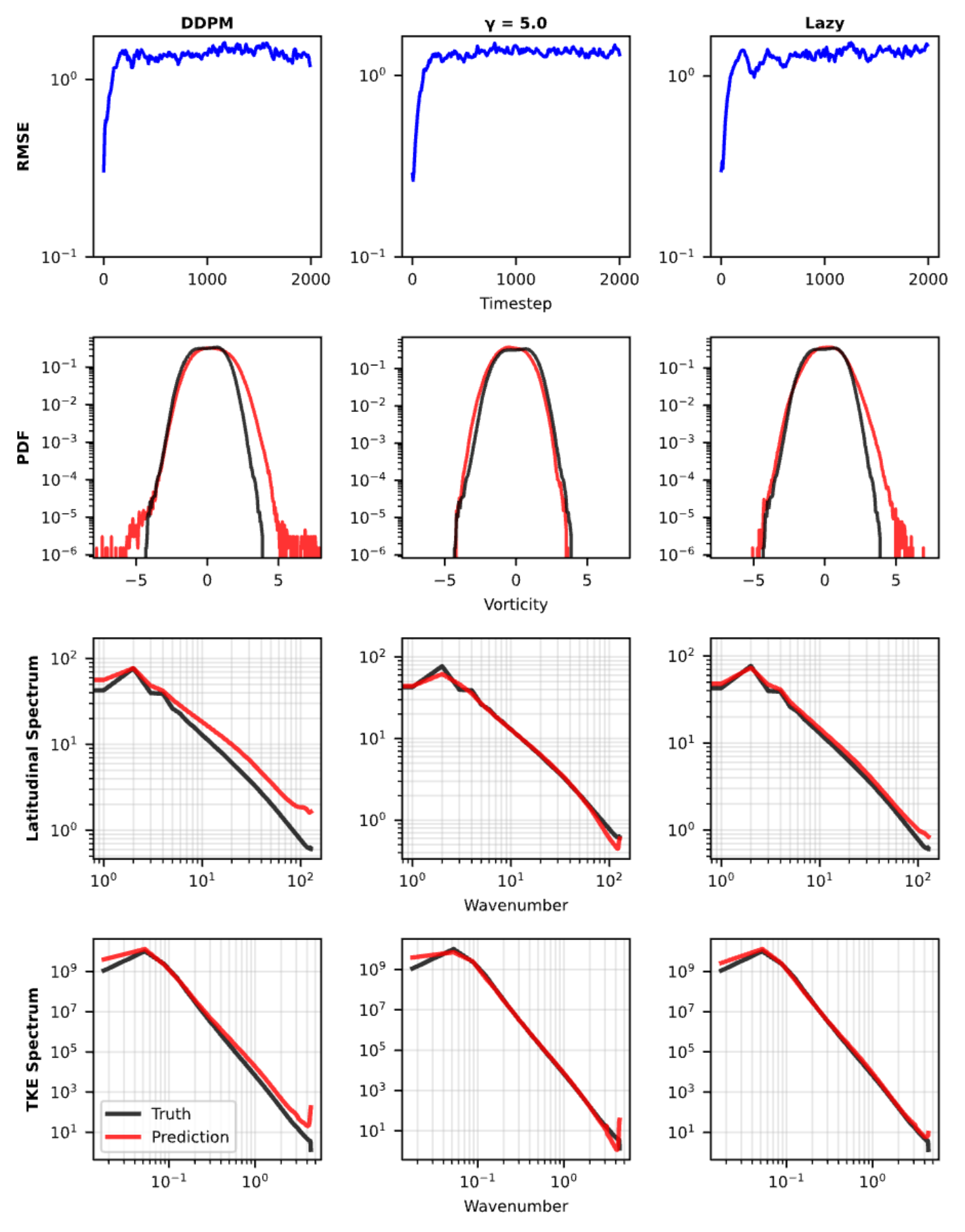}
    \caption{Quantitative evaluation metrics over 2000-timestep autoregressive rollouts for linear ($\gamma = 1$ ) DDPM, $\gamma = 5.0$ power schedule, and Lazy retraining ($\gamma = 5.0$). Row 1 shows Relative RMSE at each timestep. Row 2 Shows the total histogram of of vorticity values over all autoregression times. Row 3 shows the latitudinal spectrum averaged over the rollout. Row 4 shows the turbulent kinetic energy (TKE) spectra averaged over the whole rollout.}
    \label{fig:metrics}
\end{figure}

Standard accuracy metrics obscure these differences. As shown in Fig.~\ref{fig:metrics}, relative RMSE remains similar across models during the 2000-step rollout: DDPM achieves 1.366, the $\gamma=5.0$ model 1.292, and lazy diffusion 1.274. This highlights a key limitation: RMSE penalizes magnitude errors but is blind to structural drift and spectral mode misallocation.

More physically meaningful metrics expose the differences. The vorticity distributions show that $\gamma=5.0$ closely matches the ground truth, while DDPM displays excess probability mass at extreme vorticity values and a rightward shift, corresponding to unphysical coherent structures. Quantitatively, the Earth Mover’s Distance increases from 0.142 for $\gamma=5.0$ to 0.342 for DDPM, while KL divergence increases from 0.023 to 0.075.

Spectral analysis pinpoints the mechanism behind these failures: as seen in the latitudinal spectra of Fig.~\ref{fig:metrics}, DDPM severely overestimates power for wavenumbers $k>10$, with deviations exceeding an order of magnitude at high~$k$. In contrast, the $\gamma = 5.0$ model reproduces spectral slopes and amplitudes accurately up to the highest resolved scales. The relative spectral error for $\gamma = 5.0$ is 10.2\%, compared to 96.3\% for DDPM. Lazy diffusion exhibits mild degradation (22.7\%) but compensates with three orders of magnitude lower inference cost.

\begin{figure}[H]
    \centering
    \includegraphics[width=0.9\textwidth]{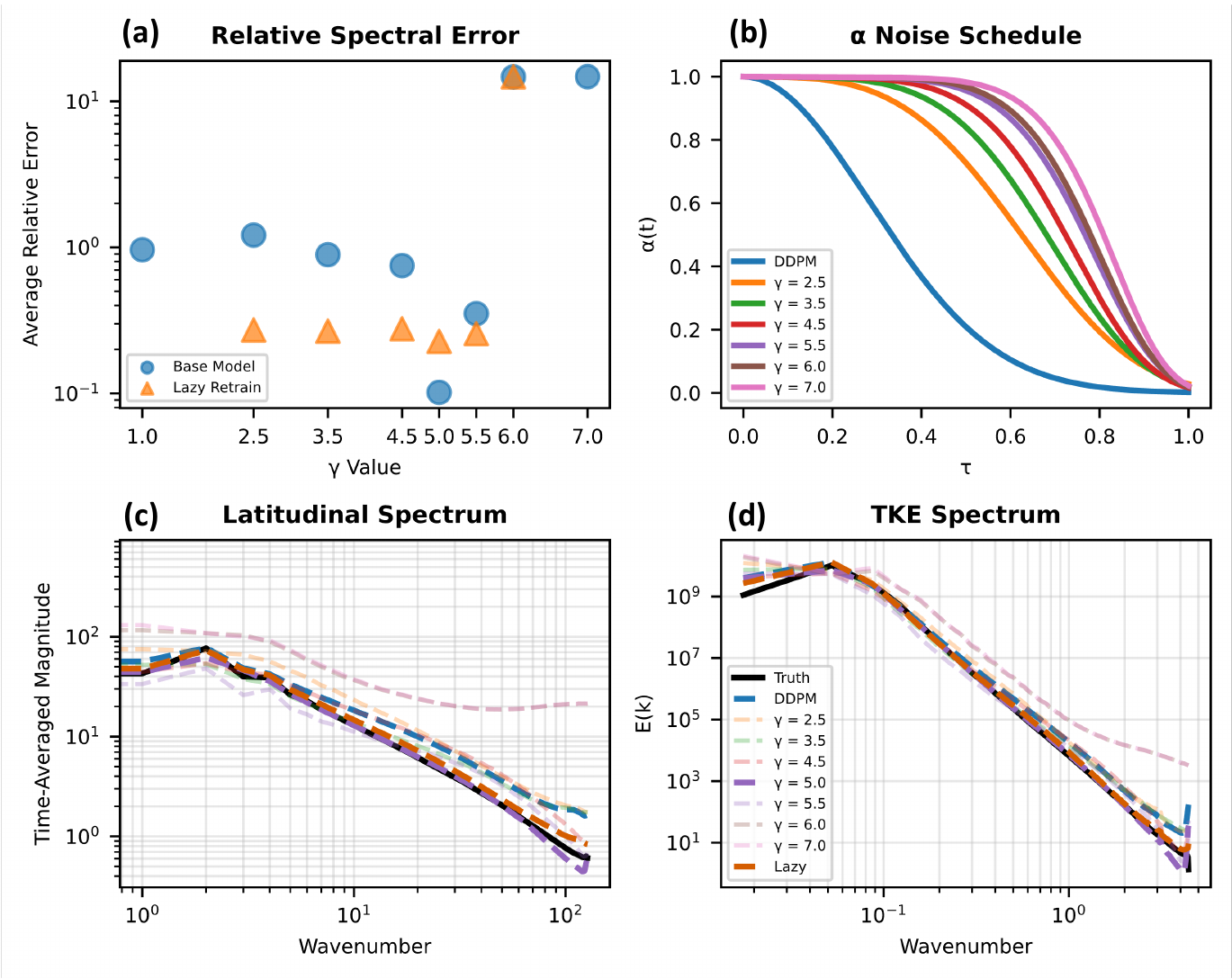}
    \caption{Power schedule analysis and spectral performance. (a) Relative spectral error as a function of power exponent $\gamma$, showing optimal performance at $\gamma = 5.0$ and catastrophic failure above $\gamma = 6.0$. Orange triangles indicate lazy diffusion retraining. (b) Signal-to-noise ratio evolution, $\alpha(\tau)$, for different power schedules, with higher $\gamma$ delaying noise injection. (c) Time-averaged latitudinal spectra of vorticity. (d) TKE spectra for all models demonstrate that appropriate power scheduling recovers high-wavenumber dynamics.}
    \label{fig:sys1_lazy_metrics}
\end{figure}

These findings are systematically reinforced by the power‐schedule sweep in Fig.~\ref{fig:sys1_lazy_metrics}. Increasing $\gamma$ from the linear schedule ($\gamma = 1$) progressively improves spectral fidelity, peaking at $\gamma = 5.0$. The improvement stems from delayed noise injection: the linear schedule reaches $\alpha(\tau) = 0.5$ at $\tau \approx 0.3$, whereas $\gamma = 5.0$ delays this transition until $\tau \approx 0.75$, preserving high‐wavenumber information for a much larger portion of the diffusion trajectory and providing significantly more reverse‐diffusion steps for fine‐scale refinement.

For $\gamma > 5.5$, a sharp training instability emerges. At $\gamma = 6.0$ and $\gamma = 7.0$, spectral error jumps by two orders of magnitude to 1470\% and 1476\%, respectively. These failures occur because excessively aggressive schedules push nearly all noise injection to the terminal portion of the forward process, where the discrete DDPM formulation cannot resolve the resulting large noise jumps. Training becomes numerically unstable without impractically small timestep discretization.

Lazy diffusion improves the performance of most base models, regardless of $\gamma$, as indicated by the orange triangles in Fig.~\ref{fig:sys1_lazy_metrics}(a). However, the best lazy variant arises from the best base model: $\gamma = 5.0$. Lazy diffusion, therefore provides a computationally efficient alternative that inherits from but cannot exceed the stability region of the underlying score model.

\begin{figure}[H]
    \centering
    \includegraphics[width=0.9\textwidth]{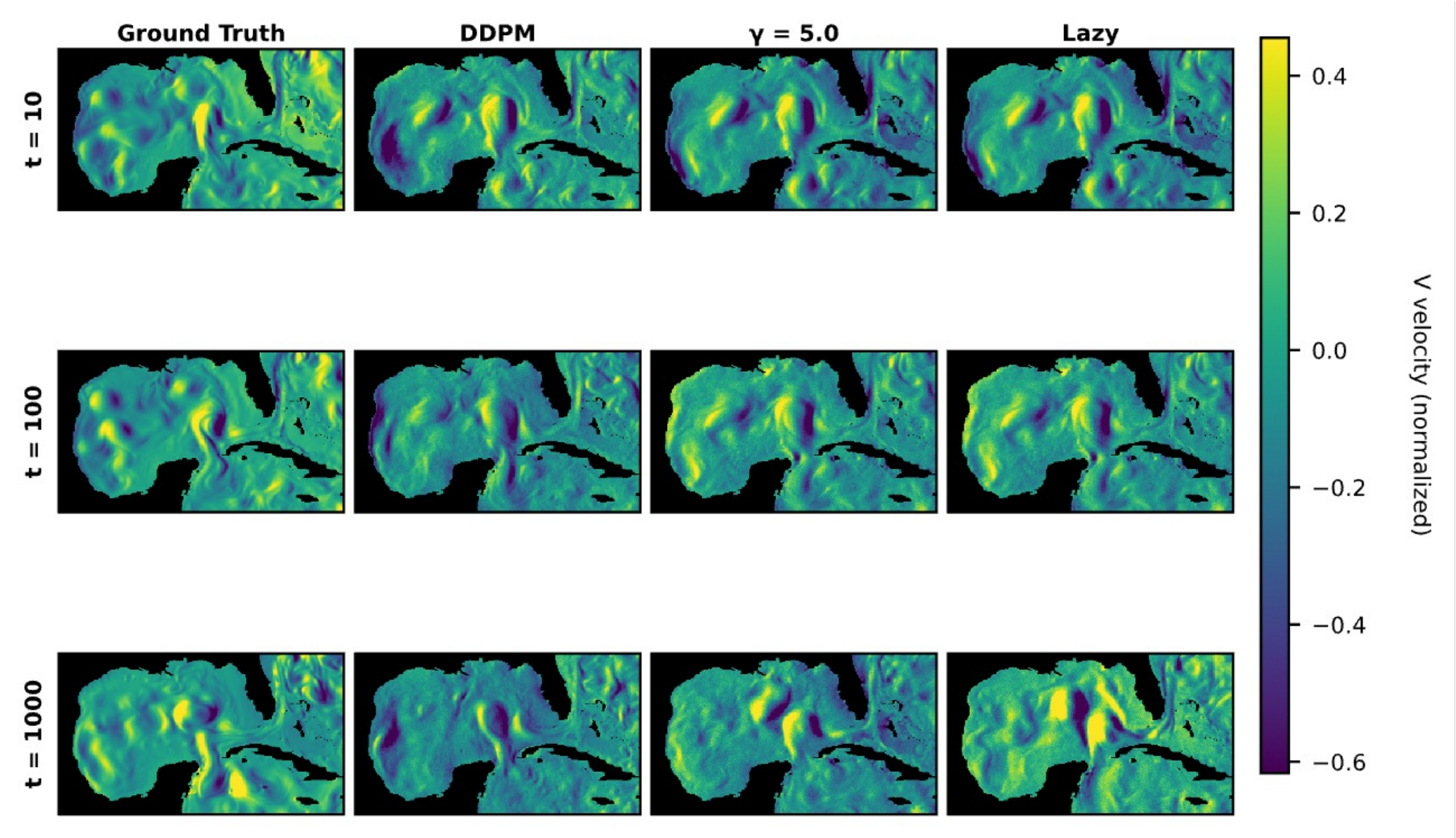}
    \caption{Autoregressive evolution of meridional ($V$) velocity from the GLORYS reanalysis dataset. Snapshots shown at $t = \{10, 500, 1000\}$ timesteps ($t=$ 1 day) for ground truth, DDPM, and the highest performing $\gamma$ $(=5.0)$ and lazy retrain. Color scale shows normalized vorticity.}
    \label{fig:sys2_sbapshot}
\end{figure}

\begin{figure}[H]
    \centering
    \includegraphics[width=0.9\textwidth]{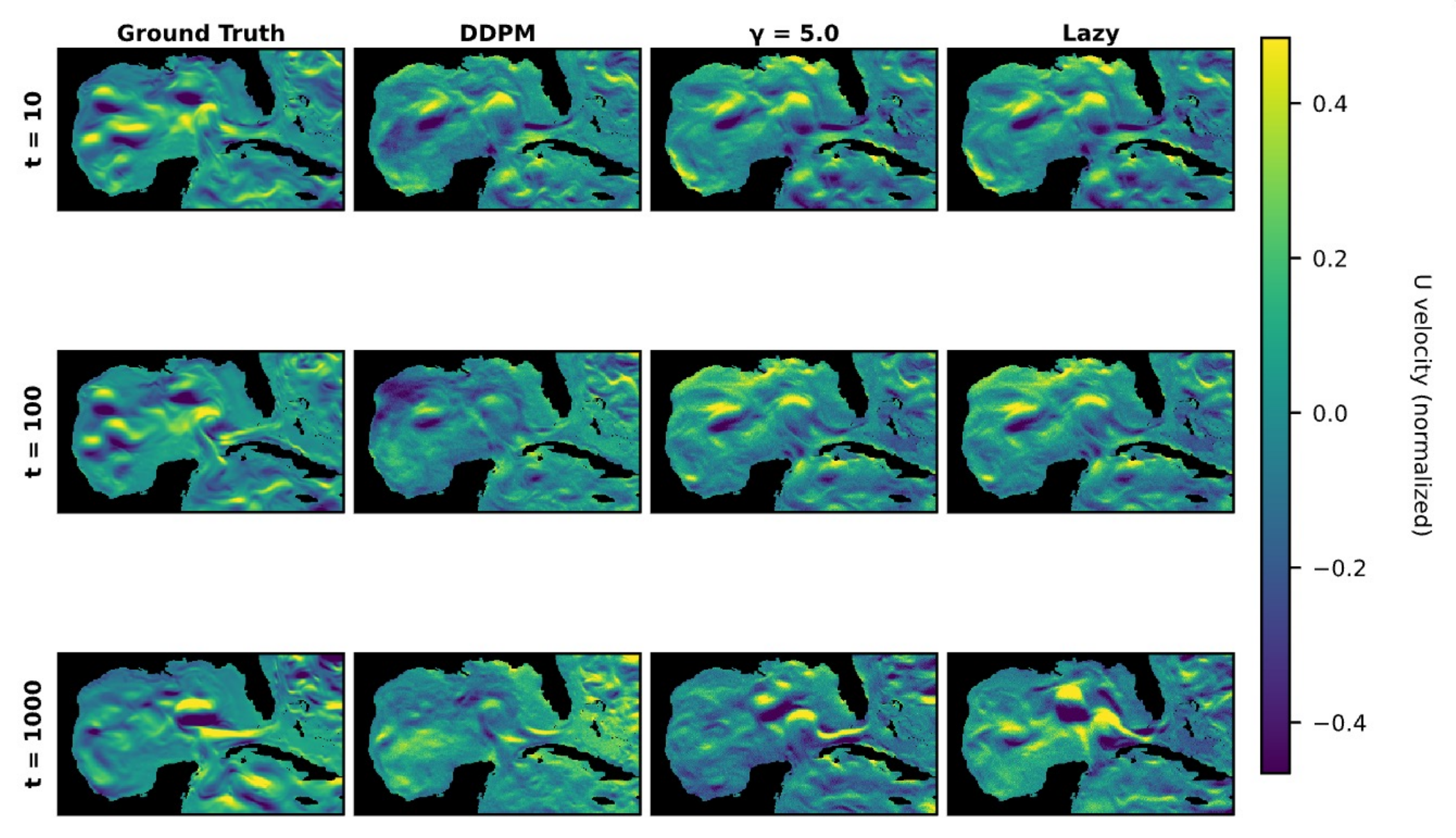}
    \caption{Autoregressive evolution of zonal velocity ($U$) from the GLORYS reanalysis dataset. Snapshots shown at $t = \{10, 500, 1000, 2000\}$ timesteps ($t=$ 1 day) for ground truth, DDPM, and the highest performing $\gamma$ $(=5.0)$ and lazy retrain. Color scale shows normalized vorticity.}
    \label{fig:sys2_snapshotU}
\end{figure}

\begin{figure}[H]
    \centering
    \includegraphics[width=0.9\textwidth]{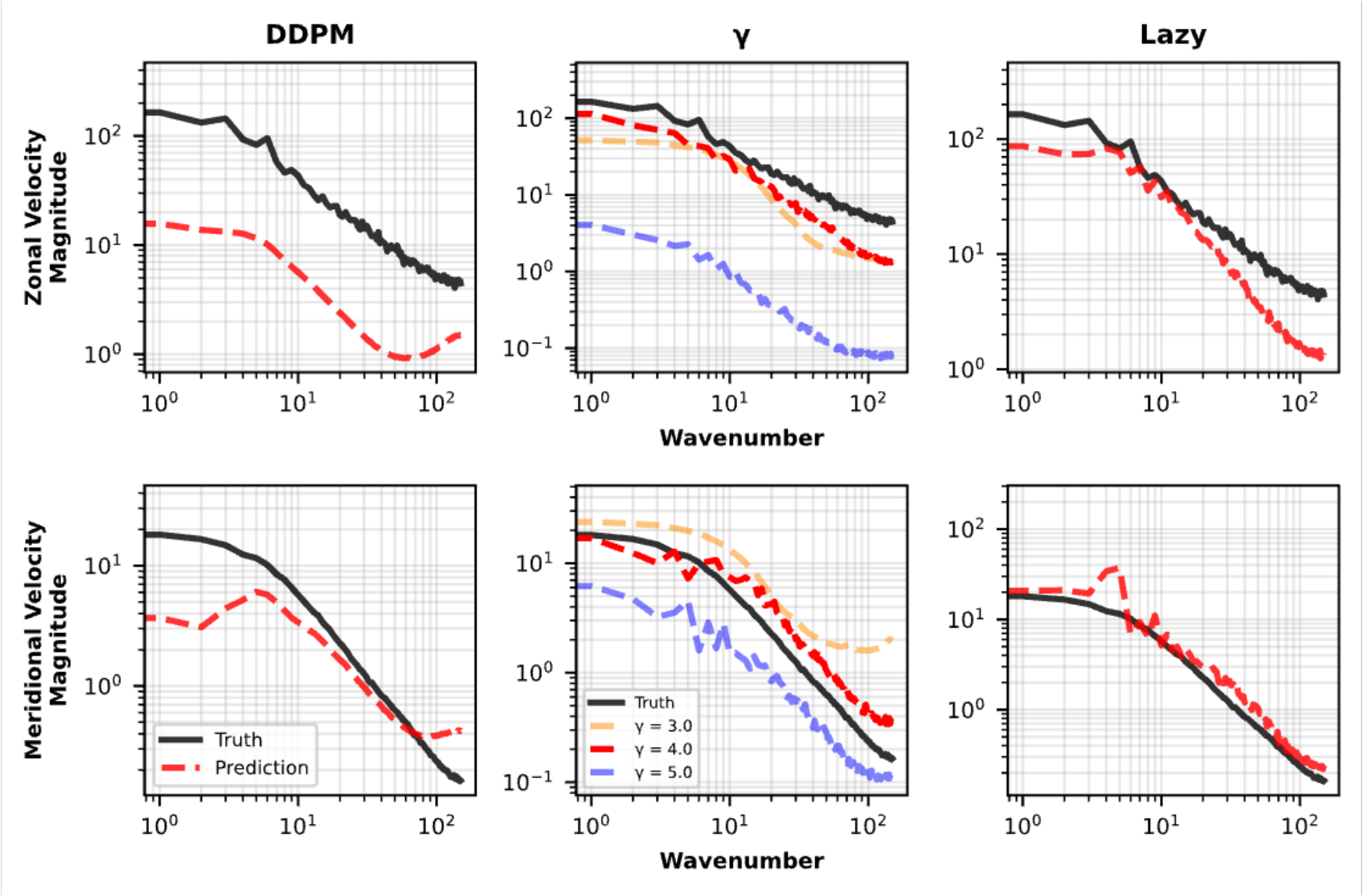}
    \caption{Quantitative evaluation metrics over 1000-timestep autoregressive rollouts for linear ($\gamma = 1 $) DDPM, $\gamma = 5.0 $ power schedule, and Lazy retraining ($\gamma = 1 $). Row 1 Shows the total histogram of Zonal velocity over all autoregression times. Row 2 Shows the total histogram of Meridional values over all autoregression times. Row 3 shows the Latitudinal spectrum averaged over the rollout. Rows 3 and 4 shows the Zonal and Meridional velocity spectra averaged over all autoregression times.}
    \label{fig:sys2_metrics}
\end{figure}

Turning to \textbf{System~2}, similar trends emerge. As shown in Figs.~\ref{fig:sys2_sbapshot}–\ref{fig:sys2_metrics}, models with the power-law noise schedule accurately preserve both zonal and meridional ocean velocities across 1000-step rollouts, whereas the linear DDPM systematically exhibits spectral energy deficits at intermediate and small scales. The $\gamma = 5.0$ model and its lazy variant reproduce the correct latitudinal, zonal, and meridional spectral slopes of the ocean reanalysis, while DDPM displays rapid drift toward overly dissipative dynamics.

Across both systems, the spectral improvements obtained through power‐law scheduling persist at all dynamically relevant scales. Both the latitudinal and TKE spectra show that $\gamma = 5.0$ and its lazy-diffusion counterpart recover the correct power‐law behavior through the inertial range and maintain high‐wavenumber structure. DDPM, by contrast, exhibits systematic energy loss beginning at intermediate wavenumbers, producing excessive dissipation in physical space.

Together, these results show that power-law noise schedules and lazy diffusion address the precise theoretical failure modes identified earlier: they counteract the wavenumber-dependent SNR collapse, redistribute training emphasis away from irreducible noise, preserve high-$k$ information deep into the diffusion horizon, and produce stable, structure-preserving, computationally efficient surrogates for turbulent flow.

\newpage

\section{Conclusion}
This work establishes a rigorous theoretical and practical foundation for fully probabilistic diffusion-based emulation of multiscale dynamical systems. By deriving a Fourier-space characterization of the DDPM forward process, we demonstrated that standard Gaussian noise schedules induce a \emph{wavenumber-dependent spectral collapse}: the signal-to-noise ratio decays monotonically in $|k|$ for power-law spectra, eliminating high-wavenumber information early in diffusion time. This analysis reveals that conventional diffusion models are structurally ill-suited for turbulent systems, whose inertial-range dynamics require persistent access to fine-scale structure. Our results show that this collapse is not merely a numerical artifact but a fundamental property of the DDPM formulation when applied to power-law-distributed physical data.

Building on these insights, we introduced \emph{power-law noise schedules} that reinterpret the corruption process as a spectral regularizer. By delaying noise injection, these schedules preserve high-wavenumber coherence deeper into the forward trajectory, thereby enabling the score network to access learnable gradients across the full spectral range. The empirical results on both 2D Kolmogorov turbulence and Gulf of Mexico ocean circulation show that an optimal exponent ($\gamma \approx 5$) yields substantial improvements in spectral fidelity, inertial-range scaling, and long-horizon stability. These improvements persist despite minimal architectural modifications, highlighting that the primary barrier to stable generative emulation lies in the physics–model mismatch introduced by the noise schedule. The associated lazy diffusion distillation further demonstrates that, once the score field is properly learned, a one-step generator can replicate the spectral and structural fidelity of long reverse-SDE trajectories at drastically reduced computational cost.

Despite these advances, several limitations remain. First, power-law scheduling introduces sensitivity to discretization: values of $\gamma$ that are too large cause near-singular noise injection late in the diffusion trajectory, leading to instability during training. This suggests that theoretical alignment between the continuous-time SDE and the discrete-time parameterization is crucial for preventing numerical artifacts. Second, the lazy diffusion framework, although computationally efficient, inherits the biases and failure modes of the base score model; it cannot compensate for deficiencies in the learned geometry of the conditional distribution. Moreover, the current models operate at moderate resolution and two-dimensional domains, leaving open questions regarding scalability to fully three-dimensional turbulence, anisotropic spectra, and coupled multiphysics systems.

Future work may address these limitations along several axes. From a theoretical standpoint, developing \emph{spectrally adaptive} noise schedules that respond dynamically to local energy content could provide a principled mechanism for stabilizing diffusion on nonstationary or inhomogeneous flows. Another promising direction lies in unifying diffusion with operator-learning frameworks, enabling models to generalize across domains, resolutions, or physical regimes. On the computational side, scaling lazy diffusion to multi-GPU and distributed environments could enable real-time probabilistic forecasting at resolutions approaching operational numerical weather prediction. Finally, extending the framework to joint state–parameter estimation or uncertainty propagation in hybrid physics–ML systems may further broaden the impact of diffusion-based generative modeling on oceanography, climate science, and beyond.

Overall, this work demonstrates that diffusion models, when redesigned to respect the spectral geometry of turbulence, can overcome long-standing stability challenges in autoregressive emulation. By coupling analytical understanding with algorithmic innovations, we provide a general blueprint for building physically aware, uncertainty-preserving generative surrogates capable of scaling to increasingly realistic dynamical systems.

\section*{Acknowledgement}
AC designed the research. AS conducted the research and developed the computational codes. Both the authors analyzed the results and prepared the manuscript. Both AS and AC were supported by the National Science Foundation (grant no. 2425667). Computational resources were provided by NSF ACCESS MTH240019, MTH250006 and NCAR CISL UCSC0008, and UCSC0009. The computational codes can be found in both the Zenodo link: 10.5281/zenodo.17807024 and on GitHub:https: //github.com/TACS-UCSC/LazyDiffusion.  

\bibliographystyle{unsrt}
\bibliography{main}

\newpage
\section*{Appendix}

\subsection{Detailed analysis of power-law sweeps for \textbf{System 1}}
Here, we we show the snapshots of evolution of \textbf{System 1} for different values of $\gamma$ used during the forward diffusion process.
\begin{figure}[H]
    \centering
    \includegraphics[width=\textwidth]{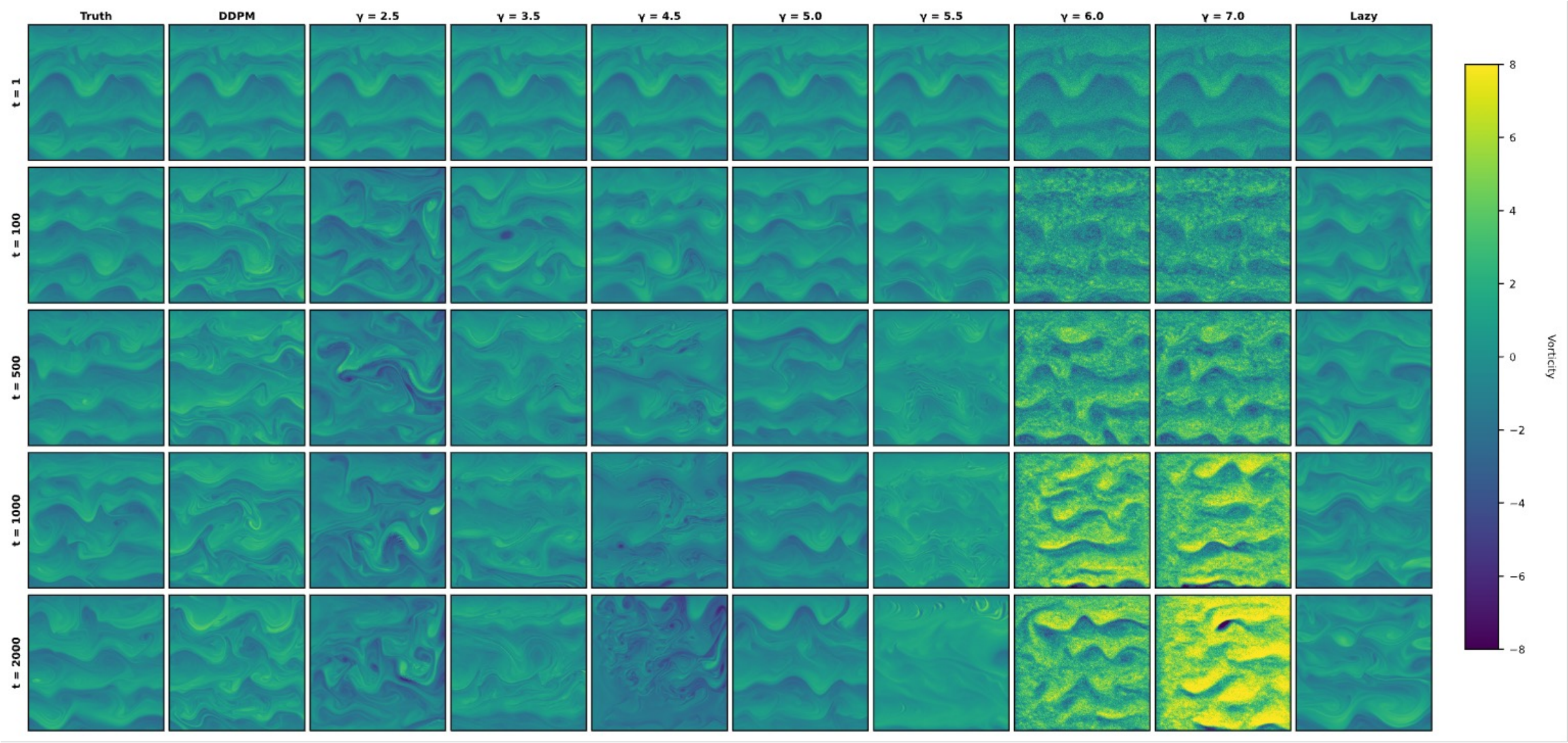}
    \caption{Autoregressive evolution of vorticity fields for 2D Kolmogorov flow. Snapshots shown at $t = \{1, 100, 500, 1000, 2000\}$ timesteps for ground truth, DDPM, a sweep of $\gamma$ values with their lazy retrains.}
    \label{fig:snapshot}
\end{figure}

\subsection{Why the score points toward a mode.} 
\label{sec:why_score_points}
Let $p_0$ denote the data density and consider a VE forward process whose marginal at diffusion time $\tau$ equals a Gaussian smoothing
\[
p_\tau(x) \;=\; (p_0 * \mathcal{N}(0,\sigma^2(\tau)\mathbf{I}))(x), \quad \sigma(\tau) > 0.
\]
Differentiating under the integral (Dominated Convergence Theorem) gives
\[
\nabla_x \log p_\tau(x)
= \frac{m_\sigma(x)-x}{\sigma^2},
\]
where
\[
m_\sigma(x) = \frac{\int y\,p_0(y)\,\phi_\sigma(x-y)\,dy}{\int p_0(y)\,\phi_\sigma(x-y)\,dy}
\]
is the Gaussian-kernel weighted mean. Thus the score direction is
\[
\operatorname{dir}(\nabla_x \log p_\tau(x)) = \operatorname{dir}(m_\sigma(x)-x).
\]
Points with $m_\sigma(x) = x$ are stationary points of the smoothed density $p_\tau$. If $p_\tau$ is locally unimodal near $x$ (or $p_0$ is a well-separated mixture at scale $\sigma(\tau)$), the gradient-ascent flow of $\log p_\tau$ from $x$ converges to the mode in its attraction basin. 

This property means that once a model has learned the score at a given noise level, moving directly toward the mode from $x_{\tau^*}$ becomes a much simpler task; the model only needs to learn an additional discretization, not the full mapping.

In practice, we found that introducing the extra training instability caused by a Variance Exploding SDE was unnecessary. Models trained using a VPSDE still exhibit stable training, likely because for normalized data the lower variance makes it easier to learn the score, which more than offsets the difficulty of learning a relatively small drift at each noise level.

\end{document}